\documentclass[preprint]{revtex4}
\usepackage{graphicx}
\usepackage{amsmath}
\usepackage{amssymb}

\begin{document}

\title{Conductance fluctuations in the localized regime}

\author{A. M. Somoza, J. Prior and M. Ortu\~no}
\affiliation{Departamento de F\'{\i}sica,
     Universidad de Murcia, Murcia 30.071, Spain}

\begin{abstract}
We have studied numerically the fluctuations of the conductance, $g$,
in two-dimensional, three-dimensional and four-dimensional disordered
non-interacting systems.
We have checked that the variance of $\ln g$ varies with
the lateral sample size as $L^{2/5}$ in three-dimensional systems,
and as a logarithm in four-dimensional systems.
The precise knowledge of the dependence of this variance with system size
allows us to test the single-parameter scaling hypothesis in
three-dimensional systems. We have also calculated the third cumulant
of the distribution of $\ln g$ in two- and three-dimensional systems,
and have found that in both cases it diverges with the exponent of 
the variance times $3/2$, remaining relevant in the large size limit.
\end{abstract}
\maketitle

\section{Introduction}

The distribution function of the conductance of disordered systems
is very well understood in the metallic regime, but poorly understood
in the localized phase.
According to the single-parameter scaling (SPS) hypothesis \cite{AA79}
the full conductance distribution
function is governed by a single parameter. It is usual to choose the ratio of the
system size $L$ to the localization length $\xi$ as the scaling parameter,
but one could alternatively choose, for example, the average of the conductance.

The validity of the SPS hypothesis has been thoroughly checked
in one--dimensional (1D) systems. In this case, it has been shown that,
all the cumulants of $\ln g$ scale linearly with
system size \cite{Ro92}. Thus,
the distribution function of $\ln g$ approaches a Gaussian
form for asymptotically long systems. In this limit, this distribution
is fully characterized by two parameters, the mean $\langle \ln g\rangle$
and the variance of $\ln g$
\begin{equation}
\sigma ^2 =\langle \ln^2 g\rangle- \langle\ln g\rangle^2\, .
\end{equation}
Both parameters are related to each other through the relation
\begin{equation}
\sigma^2 \xi/L =1\;,
\label{tau}\end{equation}
which justify SPS in 1D systems.
Here $\xi$ is the localization length, defined
in terms of the decay of the average of the logarithm of the conductance
as a function of the length of the system $L$ as
\begin{equation}
\xi= -\lim_{L \to \infty} \frac{2L}{\langle \ln g\rangle}\, .
\end{equation}
Eq.\ (\ref{tau}) was first derived  within the
so-called random phase hypothesis \cite{At80}, and it has been proven to 
hold for most models. However, it is not verified in the band tails
\cite{DL01}, in the center of the band \cite{DE03,ST03} or for very
strong disorders \cite{SP90}.

The situation in higher dimensions
is not as clear as in 1D systems. In those dimensions is far more
difficult to do analytical
calculations and numerical simulations have been limited until
recently to small sample sizes.
In the strong localization regime,
$\ln g$ was claimed to be normally distributed and the variance
was assumed to depend linearly on size in two-dimensional (2D) and 
three-dimensional (3D) systems \cite{CM87,KK92}. Recently,
it has been pointed out that this distribution is not log normal \cite{PS05b,MM05}.
Slevin, Asada and Deych \cite{SA04}
studied the variance of the Liapunov exponent
and checked the SPS hypothesis in 2D systems.

Nguyen {\it et al.} \cite{NS85} proposed a model to account for quantum
interference effects in the localized regime, where the tunneling amplitude
between two sites was calculated considering only the shortest or
forward-scattering paths. In the SPS regime the localization length
must be much larger than the lattice constant, this means that contributions from
other paths cannot be negligible. But, as the contribution of each path decay
exponentially with its distance, we can expect that some properties of
the conductance distribution (in particular the size dependence) should be dominated
by the shortest paths.
Medina and Kardar \cite{MK92,MK89} studied in detail the model. They
computed numerically the probability distribution for
tunneling and found that is approximately log normal, with its variance
increasing with distance as $r^{2/3}$ for 2D systems.
This is in contrast with the 1D case, where the variance grows linearly with
distance, and with the implicit assumptions of some works on 2D systems.

For 2D systems, we found numerically that the variance behaves as \cite{PS05}
\begin{equation}
\sigma^2=A\langle -\ln g\rangle^{\alpha}+B
\label{2D}
\end{equation}
with the exponent $\alpha$ equal to $2/3$. This is a general result that holds for
the Anderson model and Nguyen {\it et al.} (NSS) model and for different
geometries in 2D systems. The constants $A$ and $B$ are model or geometry dependent.
The precise knowledge of the dependence of $\sigma^2$ with $\langle -\ln g\rangle$
made much easier the numerical verification of the SPS hypothesis. We checked this
hypothesis for the Anderson model \cite{PS05}. As we will see, the NSS model does
not verify the SPS hypothesis.
From a numerical point of view it is more convenient to use $\langle -\ln g\rangle$ 
than $L/\xi$ as the scaling variable, since one does not have to calculate 
the localization length. It also facilitates a possible comparison with experiments.

The conductance distribution in the localized regime in 3D systems
has been studied recently by Marko\v{s} {\it et al.} \cite{MM04}.
They showed that this distribution is not log normal
and tried to fit it with an analytical model \cite{MM04}.  
The relationship between the different moments of the distribution has not been 
deeply considered. The applicability of the SPS hypothesis in the critical
regime of the Anderson model has also been verified \cite{r1,r2}.

Our goal in this paper is to find the dependence
of the variance of $\ln g$  with its average for dimensions higher than two 
in the strongly localized regime. In particular, we will investigate if Eq.\ 
(\ref{2D}) is valid in these dimensions, with a dimension dependent exponent $\alpha$.
We will also study the third central moment in 3D systems.

In the next section, we describe the two models we have used in our calculations. 
In section 3, we present the results for the dependence of the variance of 
$\ln g$ in 3D systems. We will see that our results are consistent with the SPS 
hypothesis. In section 4, we extent the previous results to four-dimensional
(4D) systems and discuss the independent path approximation. 
In section 5, we show the behavior of 
the third cumulant, equal to the third central moment, and the skewness. 
We also comment about higher order cumulants.
Finally, we discuss the results and extract some conclusions.

\section{Model}

We have studied numerically the Anderson model for 3D samples and
NSS model for 3D and 4D samples.
For the Anderson model, we consider cubic samples of size
$L\times L\times L$ described by the standard Anderson Hamiltonian
\begin{eqnarray}
H = \sum_{i} \epsilon_{i}a_{i}^{\dagger}a_{i}+ t\sum_{i,j}
a_{j}^{\dagger} a_{i}+ {\rm h.c.} \;, \label{hamil}
\end{eqnarray}
where the operator $a_{i}^{\dagger}$ ($a_{i}$) creates (destroys)
an electron at site $i$ of a cubic lattice and $\epsilon_{i}$ is
the energy of this site chosen randomly between $(-W/2, W/2)$ with
uniform probability. The double sum runs over nearest neighbors.
The hopping matrix element $t$ is taken equal to $-1$, which set
the energy scale, and the lattice constant equal to 1, setting the
length scale. All calculations with the Anderson model are done at an energy
equal to 0.01, to avoid the center of the band.

We have calculated the zero temperature conductance $g$ from the Green functions.
The conductance $G$ is proportional to the transmission
coefficient $T$ between two semi--infinite leads attached at opposite
sides of the sample
\begin{equation}
g= \frac{2e^2}{h}T
\label{res}\end{equation}
where the factor of 2 comes from spin. From now on, we will measure
the conductance in units of $2e^2/h$.
The transmission coefficient can be obtained from the Green function, 
which can be calculated propagating layer by layer with the recursive 
Green function method \cite{M85,V98}. 
This drastically reduced the computational effort.
Instead of inverting an $L^3\times L^3$ matrix, we just have to invert
$L$ times $L^2\times L^2$ matrices.
With the iterative method we can easily solve
cubic samples with lateral dimension from $L=8$ to $35$.
We have considered ranges of disorder $W$ from 20 to 30.
The number of different realizations of the disorder employed is of 2000
for most samples.
The leads serve to obtain the conductivity from the transmission formula 
in a way well controlled theoretically and close to the experimental situation.
We have considered wide leads with the same section as the samples, which 
are represented by the same hamiltonian as the system, Eq.\ (\ref{hamil}), 
but without diagonal disorder.
We use cyclic periodic boundary conditions in the direction
perpendicular to the leads.

We have also studied the NSS model \cite{NS85} and Medina and
Kardar \cite{MK92,MK89} in 3D cubic samples and 4D hypercubic samples.
In this model, one considers an Anderson hamiltonian, Eq.\ (\ref{hamil}),
in which the diagonal disorder can only take two values
$W/2$ and $-W/2$. One concentrates on the transmission amplitude between two points
in opposite corners of the sample and assumes that the quantum trajectories
joining these two points have to follow one of the (many) shortest possible paths.
The transmission at zero energy is equal to \cite{MK92}
\begin{eqnarray}
T=\left(\frac{2t}{W}\right)^{2l} J^2(l) \;, \label{mk}
\end{eqnarray}
where the transmission amplitude $J(l)$ is given by the sum over all
the directed or forward-scattering paths
\begin{eqnarray}
J(l)=\sum_{\Gamma}^{\rm directed}J_{\Gamma} \;, \label{mk2}
\end{eqnarray}
The contribution of each path, $J_\Gamma$, is the product of the 
signs of the disorder along the path $\Gamma$
and $l$ is the length of the paths. The system size $L$ is proportional
to the path length $l$, with a proportionality constant of the order of unity.
Considering only directed paths is well justified in the strongly localized regime,
where the contribution of each trajectory is exponentially small in its length.
The variance of $\ln g$ is entirely determined by $J^2(l)$ and so it depends 
on $L$, but not on $W$ in this model. To quantify the magnitude of the
fluctuations, it is convenient to do it in terms of the length $L$
in the NSS model. $\langle -\ln g\rangle$ is, of course, proportional to $L$, 
but the constant of proportionality depends on the disorder $W$.

The assumption of directed paths facilitates the computational problem 
and makes feasible to handle system sizes much larger than with the Anderson hamiltonian.
The sum over the directed path can be obtained propagating layer by layer the weight of the
trajectories starting at the initial position and passing through a certain point \cite{MK92}.
In order to maximize interference effects, it is interesting to 
consider the initial and final sites at opposite corners of the sample.  
To simplify the programming complexity of the problem in 3D systems,
we have considered a  BCC lattice with the vector joining the
two terminal points along the direction (1,0,0). In this way we
can propagate layer by layer, with each
layer being a piece of a square lattice. For 4D systems we consider 
an "hyper-BCC" lattice such that the layers that we handle are pieces 
of a simple cubic lattice.
We have calculated 3D samples with path lengths up to $l=400$
and 4D samples with $l$ up to 130.
We have averaged over $2\cdot 10^6$ realizations of the disorder in
3D systems and over $6\cdot 10^6$ realizations in 4D systems.

\section{Variance of three-dimensional systems}

The aim of this section is to find if an expression of the form
(\ref{2D}) also holds for 3D systems  with an exponent
characteristic of this dimension.
To verify this law and to determine numerically this exponent in 3D systems we first use
NSS model and represent $\sigma^2$ versus $L$
on a double-logarithmic scale. Once we have an estimate of the exponent, which
in our case turn out to be $\alpha=2/5$,
it is more demanding to represent $\sigma^2$ as a function of
$L^\alpha$ on linear scales.
This is what we do in Fig.\ 1 for NSS model in 3D systems. 
Each dot corresponds to a different system size. We see that the data follow
an excellent straight line and so we conclude that, at least in this model,
the variance of $\ln g$ verify the law
\begin{equation}
\sigma^2=A\langle -\ln g\rangle^{2/5}+B\, ,
\label{law}
\end{equation}
where $A$ and $B$ are model and/or geometry dependent constants.
We have taken into account that $\langle -\ln g\rangle$ is proportional to $L$.

\begin{figure}[htb]
\includegraphics[width=.48\textwidth]{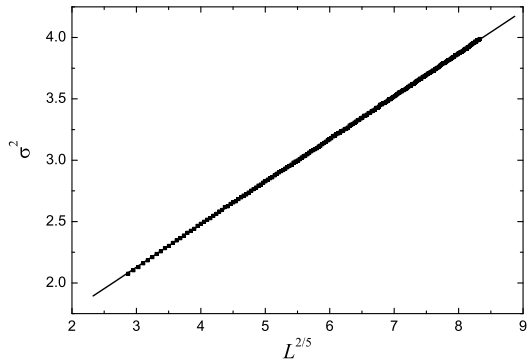}
\caption{$\sigma^2$ as a function of $L^{2/5}$ 
for NSS model in 3D systems.}
\label{fig1}
\end{figure}

\begin{figure}[htb]
\includegraphics[width=.48\textwidth]{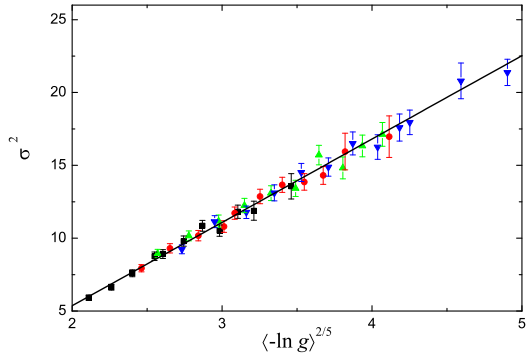}
\caption{(Color online) $\sigma^2$ versus $\langle -\ln g\rangle^{2/5}$ for the
Anderson model in 3D systems. Each symbol corresponds to a different disorder:
20 (squares), 25 (solid dots), 27 (up triangles) and 30 (down triangles).}
\label{fig2}
\end{figure}

NSS model is very convenient, because allows us to handle 
large sample size, but it involves certain approximations and does not 
satisfy the SPS hypothesis, since the variance $\sigma^2$ does not depend 
on disorder, while $\langle -\ln g\rangle$
depends on both disorder and system size.
Moreover, the conductivity of this model would correspond to a system with 
microscopically narrow leads attached to
 two points of the system, while in real systems wide leads of the order of
 the sample size are used. So, it is desirable to
see if the previous results also apply to the Anderson model with wide leads.
In Fig.\ \ref{fig2} we plot $\sigma^2$ as a function of
$\langle -\ln g\rangle^{2/5}$ for the Anderson model with wide leads in 3D 
systems for different values of the disorder
$W=20$ (squares), 25 (solid dots), 27 (up triangles) and 30 (down
triangles). We see that the data again fit a straight line quite well. 
The straight line in Fig.\ \ref{fig2} is a least square fit to the data and is given
by 
\begin{equation}
\sigma^2=(5.71\pm 0.12)\langle -\ln g\rangle^{2/5}-6.0\pm 0.3.
\label{lawc}
\end{equation}
Care must be taken when extracting the power law behavior
from a double logarithmic plot of $\sigma^2$ versus $\langle -\ln g\rangle$,
due to the presence of the constant term, which in 3D systems is quite noticeable.
Obviously, the previous behavior, Eq.\ (\ref{lawc}), must break down since the 
behavior in the metallic part must be very different and, in any case, the
variance cannot be negative. 
The critical point for the metal-insulator transition has been deeply studied
and, for periodic boundary conditions, it corresponds to $\langle -\ln g\rangle=1.280$
(in our units) for the largest sample size employed in Ref.\ \onlinecite{SW99}
and to 1.329 for the largest size in Ref.\ \onlinecite{BH01}. The variance at 
the critical point is in both works equal to 1.09.
The crossing of our fitted behavior for the variance, Eq.\ (\ref{lawc}), with
this value of 1.09 gives us an estimate of the average of $-\ln g$ at the critical
point of 1.24 in very good agreement with the values obtained in Refs.\ 
\onlinecite{SW99,BH01}. We expect Eq.\ (\ref{lawc}) to be pretty well satisfied
in the entire localized regime down to the critical point. 

A theoretical model for the conductance distribution in 3D systems in the
localized regime was proposed by Marko\v{s} {\it et al.} \cite{MM04}. These authors 
calculated numerically the dependence of the variance of $\ln g$ with its average 
and argued that this was compatible with their analytical model.
In the strongly localized regime, this model predicts a linear dependence 
($\sigma^2 \approx \langle \ln g\rangle$), in disagreement with our results,
which expand a larger range of values of $\langle \ln g\rangle$.

The data for the Anderson model in Fig.\ \ref{fig2} for the different values 
of the disorder overlap within the error bars in a single line.
The good overlap of the data is a strong support of the validity of
the SPS hypothesis, since it shows that the variance $\sigma^2$ only depends on 
the mean of $-\ln g$ and not on $W$ and $L$ separately.

The exponent $\alpha$ that relates the variance of $\ln g$ and 
$\langle -\ln g\rangle$ in Eq.\ (\ref{2D})
is 1 in 1D systems, ${2/3}$ in 2D systems and ${2/5}$ in 3D systems. 
These three results can be summarized in the following heuristic law
\begin{equation}
\alpha=\frac{2}{2^{d-1}+1}\, ,
\label{law3}
\end{equation}
where $d$ is the dimensionality of the system.

\section{Variance of four-dimensional systems}

An interesting question is to know if  Eq.\ (\ref{2D})
also applies to 4D systems. It is quite difficult to check this with the Anderson
model, but no so with NSS model. We have been able to calculate
lateral sizes up to 120 and average over $2\cdot 10^6$ samples.
The results
seem to indicate that the variance of $\ln g$ goes as the logarithm of $L$.
In Fig.\ 3 we plot $\sigma^2$ as a function of $L$ in a
logarithmic scale.
We note that the data can be fitted fairly well by a straight line.
In the inset of Fig.\ 3 we represent the same data as in the main part of the 
figure as a function of $L^{2/9}$ on linear scales. The straight line is a linear
least square fit of the data for large sizes only. The exponent $2/9$ is the
one predicted by Eq.\ (\ref{law3}) for 4D systems. Other extrapolation
schemes of the power law exponent to four dimensions will likely produce
values larger than this one. It is easy to appreciate in the 
inset of Fig.\ 3 than the logarithmic fit (main part of the figure) is better
than this power law fit.  A power law behaviour cannot 
be ruled out if finite size effects were relevant, although in 2D and 3D 
systems these effects are negligible.  

\begin{figure}[htb]
\includegraphics[width=.48\textwidth]{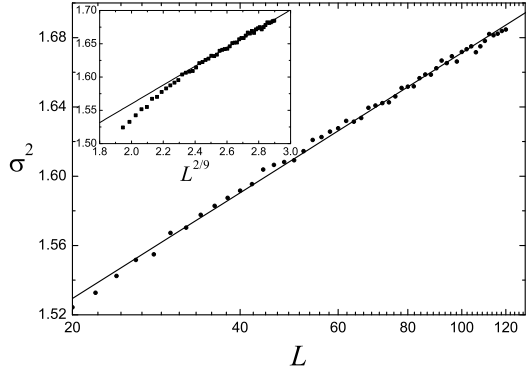}
\caption{$\sigma^2$ as a function of $L$ in a logarithmic
scale for NSS model in 4D systems. Inset: the same data as in the main part of the 
figure as a function of $L^{2/9}$.}
\label{fig3}
\end{figure}

In order to explain the results in 2D and 3D systems, one would have to
take into account the correlations between the different trajectories ending in a given
point. An approach has been developed along these lines for 2D systems \cite{MK92}.
These authors were interested in calculating $\langle J^{2n} \rangle$, where $J$ is
given by Eq.\ (\ref{mk2}) (odd powers of $J$ cancel by symmetry). 
They mapped approximately this problem to a quantum system of $n$ interacting particles
in one dimension less  than the real dimension, $d'=d-1$. The interaction between particles takes 
approximately into account the effect on intersections between different paths.
As noted by Medina and Kardar, for dimension $d>3$ ($d'>2$ for the quantum system) there is 
a possible phase transition, depending on the amount of attraction. For large enough attraction
the $n$ particles are bounded (intersections between paths are important), otherwise
the attraction becomes asymptotically irrelevant and the behaviour is like free particles
(intersections between paths are irrelevant). Obviously the expected relationship between 
mean and variance of $\ln g$ must change if we cross this phase transition. 
In the case that Eq.\ (\ref{law3}) could be applied also to 4D systems it should correspond
to the region of bound states. The NSS model does not have free parameters to control 
the effect of intersections and it seems that for 4D systems we  have already crossed the 
phase transition. It might be interesting  to test Eq.\ (\ref{law3}) in the region of bound states,
but this requires a new different model.

In order to explain the results for 4D, we may assume that the intersections between paths 
are asymptotically irrelevant and apply the independent path approach \cite{EH86}.
In this approximation one assumes that the
contributions $J_\Gamma$ of the different paths in Eq.\ (\ref{mk2})
are uncorrelated.
Then the distribution of $J(L)$ tends to a gaussian with zero mean, by symmetry,
and a variance proportional to the number of directed paths, which grows
exponentially with $L$. The corresponding distribution of
$\ln g$, which under these circumstances is the natural variable,
presents a mean linearly dependent on $L$ and a constant variance. Thus,
this approximation predicts $\alpha=0$.
Our results for 4D systems agree with this prediction, although with
logarithmic corrections. These  might be due to deviations from the
central limit and we think that they may persist at any finite dimension.

\section{Third and fourth cumulants}

We have previously shown that in 2D systems, unlike in 1D systems, 
the distribution of  $\ln g$ does not tend to a gaussian in the strongly 
localized regime, since the skewness tends to a constant, different from zero, 
in the limit of $\langle-\ln g\rangle$ going to infinity \cite{PS05b}. 
The skewness of $\ln g$ is defined as
\begin{equation}
{\rm Sk}=\frac{\langle (\ln g -\langle\ln g\rangle )^3\rangle}{\sigma^{3/2}},
\label{ske}\end{equation}
where the numerator is equal to the third cumulant or central
moment $\kappa_3$. The skewness tends to a finite constant if the
third cumulant scales as $\langle -\ln g\rangle^{3\alpha/2}$.

We analyse the behaviour of the third cumulant 
as a function of $\langle -\ln g\rangle$ or $L$ in 3D systems.
In figure 4 we show this third cumulant  for NSS
model versus $L^{3/5}$.
The exponent $3/5$ is the one for which the skewness takes to a finite value in 
3D systems, given that the variance grows as $L^{2/5}$. As the data follow a
straight line, we deduce that the distribution tends to a constant, non gaussian shape.

\begin{figure}[htb]
\includegraphics[width=0.48\textwidth]{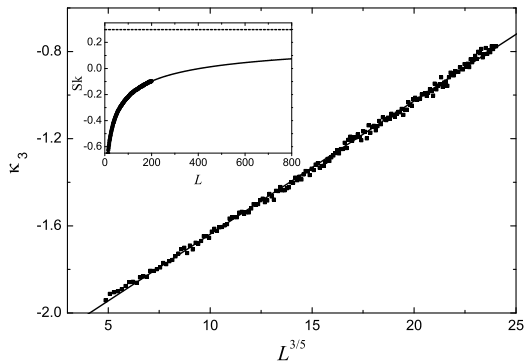}
\caption{Third cumulant of $\ln g$ versus
$L^{3/5}$ for the model of NSS
in 3D systems.
The corresponding skewness is represented in the inset.}
\label{fig4}
\end{figure}

In the inset of figure 4 we plot the skewness of the distribution of
$\ln g$ as a function of $L$ for the same data as in the 
main part of the figure. The horizontal line is our estimate of the skewness
in the macroscopic limit, equal to 0.3, obtained from the slope of the 
data in the main part of the figure.  The continuous function is obtained from the linear fits
in figures (\ref{fig1}) and (\ref{fig4}):
\[
{\rm Sk}=\frac{A L^{3/5}+B }{\left(A' L^{2/5}+B'\right)^{3/2}},
\]
where $A,B,A'$ and $B'$ are the values obtained in the corresponding fits.
We see that finite size effects are very large for the skewness due to
the constant terms appearing in the dependence of the variance and (mainly)
the third moment with $L^{2/5}$ and $L^{3/5}$, respectively.

The results for the third cumulant in the Anderson model in 3D systems are similar 
to those of NSS model. The results are consistent with
Eq.\ (\ref{cumu}), although the error bars are large.
The skewness tends to a constant value. Our estimate of this value is roughly
0.7, slightly larger than the maximum value obtained in Ref.\ \onlinecite{MM04}.

The theorem of Marcienkiewicz \cite{teo} shows that the cumulant generating
function cannot be a polynomial of degree greater than 2, that is, either
all but the first two cumulants vanish or there are an infinite number of 
nonvanishing cumulants. 
In the asymptotic limit the cumulants diverge in general, but we assume
that there is a well defined distribution after an appropriate standarization
procedure in terms of the variance.
Then, Marcienkiewicz theorem imposes a strong condition on the asymptotic behavior of
the cumulants. We are left with the following possible scenarios.
The cumulant of order $n$, $\kappa_n$, for $n>2$ grows as 
\begin{equation}
\kappa_n\propto\langle -\ln g\rangle^{n\alpha/2}
\label{cumu}\end{equation} 
where as before $\alpha$ is the exponent characterizing the variance, 
Eq.\ (\ref{2D}). In this case, the skewness and similar higher order quantities
tend to constants different from zero. The other possibility is that all the
$\kappa_n$ with $n>2$ grow with an exponent smaller than $n\alpha/2$. Then the 
distribution of $\ln g$ tends to a gaussian. The third possibility, that the 
exponent characterizing $\kappa_n$ is larger than $n\alpha/2$, would 
imply the use of this cumulant in the standarization procedure and the
normalized variance would tend to zero. We assume that this is not possible
and we are left with the first two scenario: either all higher order moments grow
like in Eq.\ (\ref{cumu}) or the distribution tends to a gaussian.

We have seen that, unlike in 1D systems, 2D and 3D systems belong to the first 
scenario, since the skewness tends to a constant, and so  we expect that 
 all higher moments verify Eq.\ (\ref{cumu}).
We have verified this result for the fourth cumulant for the NSS model
in 2D and 3D systems. 
In Fig.\ 5 we represent the fourth cumulant of the distribution of $\ln g$ 
as a function of $L^{4/3}$ for the model of NSS
in 2D systems. We note that, as expected, the data fit a straight line fairly well.

\begin{figure}[htb]
\includegraphics[width=0.48\textwidth]{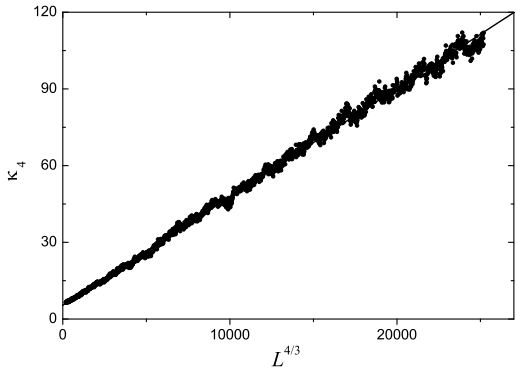}
\caption{Fourth cumulant of the distribution of $\langle -\ln g\rangle$ versus
$L^{4/3}$ for the model of NSS
in 2D systems.}
\label{fig5}
\end{figure}

\section{Discussion and conclusions}

In Table 1 we summarize the results for the exponent with which the different
cumulants scale with $\langle -\ln g\rangle$ as a function of the dimensionality
of the system.
The second column corresponds to the variance, the third and the fourth columns 
to the third and fourth cumulants, respectively.
The last column is our prediction for the $n$-th cumulant taking into account our results
for the variance and the theorem of Marcienkiewicz.
The exponent for the fourth cumulant in 2D systems has only been 
obtained with the NSS model. The law for the third cumulant in 3D systems has been 
checked for the Anderson and the NSS models. The results for the NSS model are 
conclusive, while the results for the Anderson model are compatible, but not conclusive
with the corresponding exponent.

\begin{table}
\caption{Exponents relating the cumulants with $\langle -\ln g\rangle$ for
dimensions 1, 2, 3 and 4.}
\begin{ruledtabular}
\begin{tabular}{ccccc}
Dimension&$\alpha_2$&$\alpha_3$&$\alpha_4$&$\alpha_n$\\
\noalign{\hrule}
1&1&1&1&1\\
2&2/3&1&$4/3$\footnote {Exponent obtained with the NSS model only.}&$n/3$\\
3&2/5&$3/5$\footnote {Exponent obtained with the NSS model and compatible 
with the Anderson model.}&?&$n/5$\\
4&0&&&\\
\end{tabular}
\end{ruledtabular}
\end{table}

In 1D systems, we know that all cumulants are proportional to $L$ and 
we are in the second scenario,
where only the first two cumulants are relevant for large sizes.
In 2D and 3D systems, we are in the first scenario and furthermore there is one 
and only one cumulant (beside the mean) proportional to $L$. This cumulant is the 
third one in 2D systems and must be the fifth in 3D systems. The exponent $\alpha$ of
the variance (and of all cumulants) is determined from the order of this cumulant.

We have shown that  the variance of $\ln g$ grows as $\langle -\ln g\rangle^{2/5}$
in 3D disordered non--interacting systems for the Anderson and NSS models.
In the Anderson model, we have checked that the SPS hypothesis is verified 
for energies in the band.
In 4D systems, this variance goes as $\ln L$ in the NSS model.
In dimensions higher than one, the distribution of $\ln g$ does not tend to
a gaussian. So, higher order cumulants are relevant. 

We have shown that the variance of $\ln g$ for dimensions higher than 1 grows more
slowly than linear with system size. Specifically the exponents $\alpha=2/3$ in 2D systems 
and $\alpha=2/5$ in 3D systems can be checked experimentally. We consider that a
proper theoretical explanation of these two values is also an important challenge.

\acknowledgements
The authors would like to acknowledge financial support from the Spanish DGI, 
project number BFM2003--04731 and a grant (JP).

\end{document}